\documentclass[aps,prl,twocolumn,preprintnumbers,superscriptaddress]{revtex4}
\usepackage{amssymb}
\usepackage{bm}
\usepackage{amsmath}
\usepackage[dvips]{graphicx}
\usepackage{color}

\begin{document}

\title{Why phase diagrams of different underdoped cuprates are
remarkably different?
Disorder versus bilayer.}

\author{O. P. Sushkov}
\affiliation{School of Physics, University of New South Wales, Sydney 2052, Australia}

\begin{abstract}
Contrary to a widely accepted view the phase diagrams of La$_{2-x}$Sr$_x$CuO$_4$
and YBa$_2$Cu$_3$O$\rm _{6+y}$, in spite of similarities are remarkably different.
Both the electric conduction properties and the commensurate/incommensurate 
spin ordering properties differ dramatically.
It is argued that the role of disorder in YBCO is insignificant
while the bilayer structure is crucial.
On the other hand  in LSCO the intrinsic disorder to a large extent
drives the properties of the system.
The developed approach explains the low-temperature magnetic properties of the systems.
The most important point is the difference with respect to the 
incommensurate spin ordering, including the difference in the incommensurate pitches. 
The present analysis demonstrates that the superconductivity is intimately
related to the incommensurate spin ordering.
\end{abstract}

\date{\today}
\maketitle

\begin{figure}[ht]
\includegraphics[width=0.23\textwidth,clip]{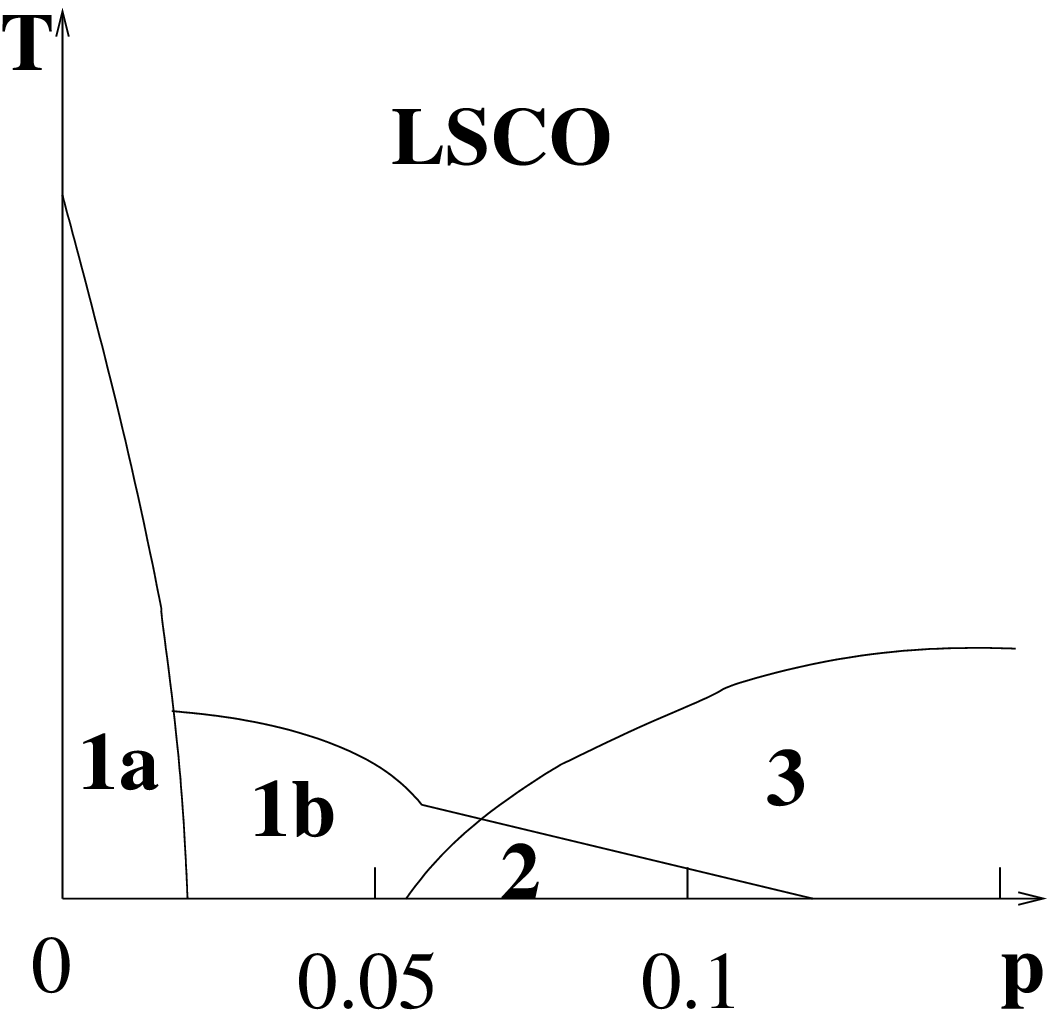}
\includegraphics[width=0.23\textwidth,clip]{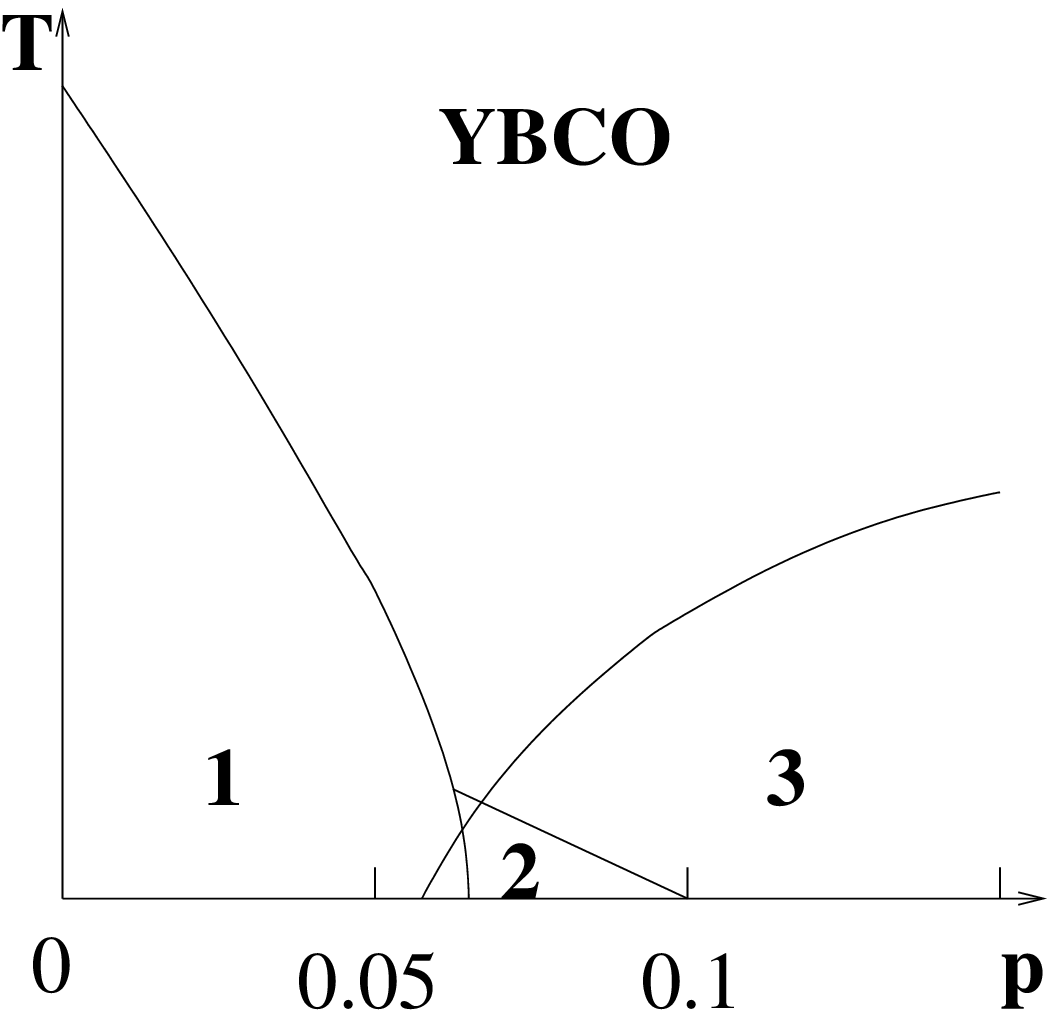}
\caption{Schematic low-doping and low-temperature phase diagrams of LSCO and YBCO.\\
LSCO: {\bf 1a.}AF order coexists with diagonal incommensurate spin structure, 
strong localization of holes.
{\bf 1b.}Diagonal incommensurate spin structure, strong localization 
of holes.
{\bf  2.}Parallel incommensurate quasistatic spin structure, 
superconductivity.
{\bf 3.}Parallel incommensurate dynamic spin structure, 
superconductivity.\\
YBCO: {\bf 1.}AF order, weak localization of holes.
{\bf  2.}Parallel incommensurate quasistatic spin structure, 
superconductivity.
{\bf 3.}Parallel incommensurate dynamic spin structure, 
superconductivity.
}
\label{lsco}
\end{figure}

 In early days of high temperature superconductivity
there was a belief that the phase diagram of La$_{2-x}$Sr$_x$CuO$_4$ (LSCO) 
represents a generic
phase diagram of cuprate superconductors.
Nowadays it has become clear that, in spite of similarities, there are
very important differences between different cuprates.
LSCO and YBa$_2$Cu$_3$O$\rm _{6+y}$ (YBSO) are
the best experimentally studied  compounds in the low doping regime.
This is why the present work addresses these compounds.
In LSCO the doping level of CuO$_2$-planes $p$
practically coincides with Sr concentration, $p\approx  x$, while in YBCO, because
of the partial filling of oxygen chains, the doping level is different
from the oxygen concentration y.
In LSCO doping gives way to superconductivity at $p > p_{sc}\approx 0.055$ and in
YBCO at $p > p_{sc} \approx 0.065$, see Fig.~\ref{lsco}. 
At first sight this indicates full similarity.
However, I will argue that the mechanisms behind $p_{sc}$ in those two compounds
are different and the closeness of the two values of $p_{sc}$ is purely accidental.
An important observation is that the normal state electrical resistivities at 
$p < p_{sc}$ are very much different.
At low temperature, $T \lesssim 100K$, and at doping below
the superconductivity threshold the in-plane resistivity
of LSCO exhibits \cite{Keimer92,ando02} the  Mott  variable-range hopping  
regime $\rho \propto \exp\{(T_0/T)^{1/3}\}$.
This indicates strong localization of holes in the N\'{e}el and the spin glass
regions of the LSCO phase diagram.
These are the regions 1a and 1b in  Fig.\ref{lsco}.
On the other hand, the in-plane resistivity in YBCO at $p < p_{sc}$ shows  only 
logarithmic dependence on temperature,  $\rho \propto \ln(C/T)$,
indicating weak-localization regime~\cite{sun,doiron}.
This is the region 1 on  the YBCO phase diagram, Fig.\ref{lsco}.
For example, at $p\approx 0.04$ the in-plane resistivity of LSCO is about 5
times larger than that of YBCO at $T=10K$, and the same ratio is about $10^3$
at $T=1K$. 
Thus, role of disorder in LSCO below the superconductivity
threshold is crucial, while in YBCO the disorder is a relatively minor issue.

The magnetic properties of the compounds are also very much different.
The three-dimensional antiferromagnetic (AF) N\'{e}el 
order in LSCO
disappears at doping $p\approx 0.02$ and gives way to the so-called
spin glass phase.
The incommensurate magnetic order has been observed at low temperature in 
neutron scattering. 
This order manifests itself as a scattering
peak shifted with respect to the AF  position. 
The incommensurate scattering has been observed even in the N\'{e}el phase
where it coexists with the commensurate one.
In the N\'{e}el phase, the incommensurability is
almost doping-independent and directed along the orthorhombic $b$ 
axis~\cite{matsuda02}. 
In the spin-glass phase, the shift is directed along the $b
$ axis, and scales linearly with doping~\cite{fujita02}. 
In the underdoped superconducting region ($0.055\lesssim
p\lesssim 0.12$), the shift still scales linearly with doping, but it is
directed along one of the crystal axes of the tetragonal lattice~\cite{yamada98}.
In YBCO the commensurate three-dimensional AF order exists up to $p \approx 0.065$, 
see Fig.~\ref{lsco}.
Moreover, there are indications that there is a  narrow window around this doping
where superconductivity and the commensurate AF order coexist~\cite{miller}.
Recently the incommensurate quasistatic spin ordering along the tetragonal 
$a^*$ direction has been observed  within the superconducting phase of 
YBCO~\cite{hinkov07a,hinkov07b} at doing $p\approx 0.085$. 
The ordering becomes fully dynamic above $p \approx 0.1$~\cite{hinkov04}. 
Last, but not least, the observed incommensurate wave vector
in YBCO at $p\approx 0.085$~\cite{hinkov07a,hinkov07b}
is of a factor two smaller than the incommensurate wave vector in LSCO at the same 
doping~\cite{yamada98} .
On the other hand, at $p\approx 0.12$ the incommensurate wave vectors in LSCO 
and YBCO are equal~\cite{yamada98,hinkov04}.

The phase diagram of underdoped LSCO has been explained 
in Refs.~\cite{sushkov05,luscher06,luscher07}.
Physics of this compound is, to a large extent, driven by disorder.
At low temperature each hole is trapped in a hydrogen-like bound state
near the corresponding Sr ion.
Each bound state creates a spiral distortion of the spin
background.
The distortion is observed in neutron scattering.
So, the state at $0.02 < p < 0.055$ is not a simple spin glass, it is a
disordered spin spiral. Both the lower and the upper boundaries of this region
are determined by the size of the bound state. The upper boundary, $p=0.055$, is
a percolation point of isolated bound states. After the percolation the 
superconductivity becomes possible, and simultaneously direction of the spin spiral
must rotate by $45^o$. The rotation is driven by the Pauli principle.
The role of disorder at $p > 0.055$ is only marginal. 
Here the spin-spiral state suggested long time ago by  Shraiman and 
Siggia~\cite{shraiman88} is  realized. Most importantly,
the state is superconducting and
the spin spiral becomes dynamic at $p > 0.12 $~\cite{milstein07}.

The present work is aimed at underdoped YBCO where, according to data on conductivity,
the role of disorder is practically insignificant. 
(This is consistent with the fact that a diagonal
spin structure has been never observed in YBCO.)
The following two issues are addressed.
1) Why does the AF order survive up to a pretty
large hole concentration $p \approx 0.06-0.07$? 
2) Why is the pitch of the incommensurate
spin order different from that in LSCO?
It will be  demonstrated that both these issues are closely related and they are
due to interlayer hopping.

The present analysis of the single CuO$_2$ -layer is based on the two-dimensional
 $t-t'-t''-J$ model at small doping.
After integrating out the high energy fluctuations one comes to the
effective low energy action of the model~\cite{milstein07}.
Importantly, the integration of the high energy fluctuations
is a fully controlled procedure, 
the small parameter justifying the procedure is the doping level, $p \ll 1$.
The effective low-energy Lagrangian is written in terms of 
the bosonic ${\vec n}$-field ($n^2=1$) that describes the staggered component of the 
copper spins,
and in terms of fermionic holons $\psi$. I use the term ``holon'' instead of ``hole'' 
because spin and charge are to large extent separated, see~\cite{milstein07}.
The holon has a pseudospin that originates from two sublattices,
so the fermionic field $\psi$ is a spinor acting on pseudospin.
Minimums of the holon dispersion are at the nodal points
$\mathbf{q}_{0}=(\pm \pi /2,\pm \pi /2)$.
So, there are holons of two types (= two flavors) corresponding to two
pockets. The dispersion in a pocket is somewhat
anisotropic, but for simplicity let us use here the isotropic approximation,
$
\epsilon \left( \mathbf{p}\right) \approx \frac{1}{2}\beta \mathbf{p}^{2}$
\ ,
where ${\bf p}={\bf q}-\mathbf{q}_{0}$.
The lattice spacing is set to be equal to unity, 3.81\thinspace \AA $\,\rightarrow $
\thinspace 1. 
All in all, the effective Lagrangian reads~\cite{milstein07}
\begin{eqnarray} 
\label{eq:LL}
{\cal L}&=&\frac{\chi_{\perp}}{2}{\dot{\vec n}}^2-
\frac{\rho_s}{2}\left({\bm \nabla}{\vec n}\right)^2\\
&+&\sum_{\alpha}\left\{ \frac{i}{2}
\left[\psi^{\dag}_{\alpha}{{\cal D}_t \psi}_{\alpha}-
{({\cal D}_t \psi_{\alpha})}^{\dag}\psi_{\alpha}\right]\right.\nonumber\\
&-&\left.\psi^{\dag}_{\alpha}\epsilon({\bf \cal P})\psi_{\alpha}  
+ \sqrt{2}g (\psi^{\dag}_{\alpha}{\vec \sigma}\psi_{\alpha})
\cdot\left[{\vec n} \times ({\bm e}_{\alpha}\cdot{\bm \nabla}){\vec n}\right]\right\} \ .
\nonumber
\end{eqnarray}
The first two terms in the Lagrangian represent the usual nonlinear 
$\sigma$ model.
The magnetic susceptibility and the spin stiffness are
$\chi_{\perp}\approx 0.53/8\approx 0.066$ and $\rho_s \approx 0.18$~\cite{SZ}.
Hereafter the antiferromagnetic exchange of the initial t-J model is set to be equal
to unity, $J\approx 130\thinspace \mbox{meV} \,\rightarrow $
\thinspace 1. 
Note that $\rho_s$ is the bare spin stiffness, therefore by definition it is
independent of doping. 
The rest of the Lagrangian in Eq.~(\ref{eq:LL}) represents the fermionic holon 
field and its interaction with the ${\vec n}$-field.
The index $\alpha=1,2$ (flavor) indicates the pocket
in which the holon resides.
The pseudospin operator is $\frac{1}{2}{\vec \sigma}$,  and 
${\bf e}_{\alpha}=(1/\sqrt{2},\pm 1/\sqrt{2})$ is a unit  vector orthogonal to the face 
of the MBZ where the holon is located. 
A very important point is that the argument of $\epsilon_{\alpha}$ in Eq.~(\ref{eq:LL}) 
is  a ``long'' (covariant) momentum,
$
{\bf {\cal P}}=-i{\bm \nabla}
+\frac{1}{2}{\vec \sigma}\cdot[{\vec n}\times{\bm \nabla}{\vec n}] \ .
$
An even more important point is that the time derivatives that stay in the
kinetic energy of the fermionic field are also ``long'' (covariant),
$
{\cal D}_t=\partial_t
+\frac{i}{2}{\vec \sigma}\cdot[{\vec n}\times{\dot{\vec n}}] \ .
$
While the semiclassical behaviour is determined by the 
Shraiman-Siggia term (the last term in (\ref{eq:LL})),
the covariant derivatives are crucial for quantum fluctuations
and in particular for stability of the system.

The effective Lagrangian (\ref{eq:LL}) is valid regardless of whether
 the ${\vec n}$-field is
static or dynamic. In other words, it does not matter if the ground state expectation
value of the staggered field is nonzero, $\langle {\vec n}\rangle\ne 0$, or zero,
$\langle {\vec n}\rangle= 0$.
The only condition for validity of (\ref{eq:LL}) is that all dynamic fluctuations
of the ${\vec n}$-field are sufficiently slow.
The typical energy of the ${\vec n}$-field dynamic fluctuations is $E_{cross}\propto p^{3/2}$,
see Ref.~\cite{milstein07},  and it must be small compared to the holon Fermi energy
$\epsilon_F \propto p$. The inequality $E_{cross} \ll \epsilon_F$ is valid up to optimal doping,
$p \approx 0.15$. So, this is the regime where (\ref{eq:LL}) is parametrically justified.
Numerical calculations within the $t-t'-t''-J$ model with physical values
of hopping matrix elements give the following values of the coupling constant
and the inverse mass, $g\approx 1$, $\beta \approx 2.2$.
On the other hand the fit of the neutron scattering data on LSCO 
gives $g\approx 1$, $\beta \approx 2.7$, which is in
good agreement with the $t-t'-t''-J$ model, see discussion in
Ref.~\cite{milstein07}.

The dimensionless parameter
\begin{equation}
\label{Omega}
\lambda=\frac{2g^2}{\pi\beta\rho_s}
\end{equation}
plays the defining role in the theory~\cite{milstein07}.
If $\lambda \leq 1$, the ground state corresponding to the Lagrangian (\ref{eq:LL})
is the usual N\'eel state and it stays collinear at  any small doping.
If $1\leq \lambda \leq 2$, the N\'eel state is unstable at arbitrarily small doping
and the ground state is a static or dynamic spin spiral.
Whether the spin spiral is static or dynamic depends on the doping level.
The pitch of the spiral is
\begin{equation}
\label{Q1}
Q=\frac{g}{\rho_s}p  \ .
\end{equation}
If $\lambda \geq 2$, the system is unstable with respect to phase separation and/or 
charge-density-wave formation and
hence the effective long-wave-length Lagrangian (\ref{eq:LL}) becomes meaningless.
By the way, the pure $t-J$ model ($t'=t''=0$) is unstable since it corresponds
to $\lambda > 2$.
Using values of $g$ and $\beta$ found from fit of experimental data,
one obtains that for LSCO $\lambda \approx 1.30$.

How the described above physics is changed in  case of YBCO? 
Due to the bilayer structure the magnon spectrum in YBCO is split into acoustic and
optic mode~\cite{Jp}. The optical gap is about 70meV. This is substantially smaller than
the maximum magnon energy $\sim 2J \sim 260\mbox{meV}$.
Therefore, the bilayer structure cannot substantially influence
values of the effective coupling constant $g$ and the inverse mass $\beta$
which are due to magnetic fluctuations with the typical energy scale $\sim 2J$.
So, one should expect that values of these parameters in YBCO are close to that in LSCO.
The holon dispersion in YBCO is split into bilayer bonding and antibonding branches
\begin{equation}
\label{eab}
\epsilon_{b,a}=\pm\frac{\Delta}{2}+\beta\frac{{\bm p}^2}{2} \ .
\end{equation}
The splitting $\Delta$ is most likely due to the hole hopping via the interlayer
oxygen chain sites. In any case both the LDA calculation~\cite{andersen95}
and the ARPES measurements~\cite{borisenko06} indicate the band splitting at nodal
points about $\Delta \sim 100\mbox{meV}$. In the present work $\Delta$ will be used as a 
fitting parameter. The splitting $\Delta$  brings additional nontrivial
physics in the system.

\begin{figure}[ht]
\includegraphics[width=0.3\textwidth,clip]{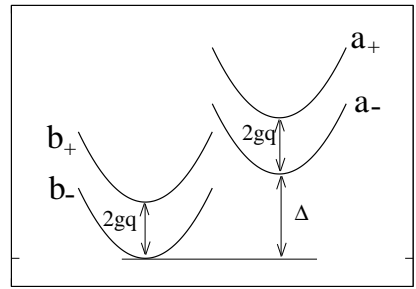}
\caption{Schematic  dispersion of a holon. b,a -corresponds to bonding and
antibonding branches, and $\pm$ corresponds to the spiral splitting.}
\label{ghh}
\end{figure}

Let us impose the coplanar spiral configuration on the system
$
{\vec n}_1=(\cos{\bf q}\cdot{\bf r},\ \sin{\bf q}\cdot{\bf r},\ 0) \ ,
$
$
{\vec n}_2=-(\cos{\bf q}\cdot{\bf r},\ \sin{\bf q}\cdot{\bf r},\ 0) \ ,
$
where ${\bm q}$ is  directed along the CuO bond
[${\bf q}\propto (1,0)$ or ${\bf q}\propto (0,1)$].
Here ${\vec n}_1$ and ${\vec n}_2$ correspond to the two layers.
Note, that ${\vec n}_1$ and ${\vec n}_2$ remain antiparralell at any given point ${\bf r}$,
hence there is no an admixture of the optic magnon to the ground state configuration. 
The single holon energy spectrum is shown 
schematically in Fig.~\ref{ghh}. There is the bonding-antibonding splitting $\Delta$,
and within each band 
there is  a splitting between different pseudospin states, $\pm gq$,
because of the spiral.
Populations of the four bands shown in Fig.~\ref{ghh} depend on 
the doping level $p$ and on the spiral wave vector $q$. 
When calculating energy, one has to remember that there are two planes.
Therefore, the elastic energy per unit area is $2\times \rho_sq^2/2$, and
the hole density per unit area is $2p$. It is convenient to define
the following characteristic concentration
\begin{equation}
\label{p0}
p_0=\frac{\Delta}{\pi\beta} \ .
\end{equation}
A straighforward calculation shows that the dependences
of the band filling and  energy on  doping $p$ and pitch $q$
are the following. Only the filled bands are noted below.
In the case $p < p_0/2$
the both $b$-bands are filled at $q < \frac{\pi\beta}{g} p$
and at $q > \frac{\pi\beta}{g} p$ only the band $b_-$ is filled. 
The energy is
\begin{equation}
\label{e1} 
\frac{E}{\rho_s}=
\left\{\begin{array}{ll}
(1-\frac{\lambda}{2})q^2+\frac{\pi\beta}{\rho_s} p^2 \ \ \ \ ,& q < \frac{\pi\beta}{g} p\\
q^2-2p\frac{g}{\rho_s}q+2\frac{\pi\beta}{\rho_s} p^2 \ , & q > \frac{\pi\beta}{g} p
\end{array}\right.
\quad .\nonumber
\end{equation}
In the case $p_0/2 < p < p_0$ the bands $b_{\pm}$ are filled at $q < q_1$,
the bands $b_{\pm}$ and $a_-$ are filled at $q_1< q < q_2$, and
$b_{-}$, $a_-$ are filled at $q > q_2$.
The energy reads
\begin{eqnarray}
\label{e2} 
\frac{E}{\rho_s}=
\left\{\begin{array}{lc}
(1-\frac{\lambda}{2})q^2+\frac{\pi\beta}{\rho_s} p^2 
\ \ \ \ \ \ \ \ \ \ \ \ \ \ \ \ \ \ \ \ \ \ ,    & q < q_1\\
\left(1-\frac{2}{3}\lambda\right)q^2+\frac{\lambda}{3}qq_1
-\frac{\lambda}{6} q_1^2+\frac{\pi\beta}{\rho_s} p^2
 \ , & q_2> q > q_1 \\
q^2-2p\frac{g}{\rho_s }q
+\frac{\pi\beta}{\rho_s} \left(p^2+pp_0-\frac{p_0^2}{4}\right) \  , & q > q_2
\end{array}\right.
\quad .\nonumber
\end{eqnarray}
In the case $p > p_0$ the bands $b_{\pm}$, $a_{\pm}$  are filled at $q < q_3$,
the bands $b_{\pm}$ and $a_-$ are filled at $q_3< q < q_2$, and
$b_{-}$, $a_-$ are filled at $q > q_2$.
The energy reads
\begin{eqnarray}
\label{e3} 
\frac{E}{\rho_s}=
\left\{\begin{array}{lc}
(1-\lambda)q^2+\frac{\pi\beta}{\rho_s} \left(\frac{p^2}{2}-\frac{p_0^2}{2}+p p_0\right)
\ \ \ \ \ ,    & q < q_3\\
\left(1-\frac{2}{3}\lambda\right)q^2-\frac{2\lambda}{3}qq_3
-\frac{2\lambda}{3} q_3^2+\frac{\pi\beta}{\rho_s} p^2
 \ , & q_2> q > q_3 \\
q^2-2p\frac{g}{\rho_s }q
+\frac{\pi\beta}{\rho_s} \left(p^2+pp_0-\frac{p_0^2}{4}\right) \ \ \ , & q > q_2
\end{array}\right.
\quad .\nonumber
\end{eqnarray}
In these formulas
$q_1= \frac{\pi\beta}{g}(p_0-p)$, \ 
$q_2= \frac{\pi\beta}{g}\left(\frac{p_0}{4}+\frac{p}{2}\right)$, and
$q_3= \frac{\pi\beta}{g}\left(\frac{p}{2}-\frac{p_0}{2}\right)$.

\begin{figure}[ht]
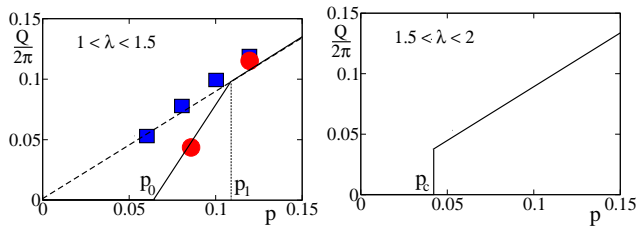

\includegraphics[width=0.23\textwidth,clip]{Q1.eps}
\includegraphics[width=0.23\textwidth,clip]{Q2.eps}
\caption{Incommensurate pitch versus doping.\\
Left: the regime $1 < \lambda < 1.5$. 
The solid line is the theory prediction for YBCO, the parameters are 
$\lambda=1.3$, $p_0=0.065$.\\
The dashed line is the theory prediction for LSCO.
The red circles represent the YBCO neutron scattering data from 
Refs.~\cite{hinkov07a,hinkov07b,hinkov04}.
The blue squares represent the LSCO neutron scattering data from 
Ref.~\cite{yamada98}.\\
Right: theoretical prediction for the pitch in the case $1.5 < \lambda < 2$.}
\label{Q12}
\end{figure}

The minimum of the energy with respect to $q$ gives the equilibrium spiral pitch $Q$
at a given doping level $p$.
The result depends on $\lambda$. For $1 < \lambda < \frac{3}{2}$ the pitch stays
zero for $p < p_0$, then for $p_0 < p < p_1$, where $p_1=0.5p_0/(\lambda-1)$, the pitch is
\begin{equation}
\label{Q2}
Q=\frac{g}{\rho_s}\frac{p-p_0}{3-2\lambda}  \ ,
\end{equation}
and finally at $p > p_1$ the pitch is given by the single layer formula (\ref{Q1}).
This behaviour is illustrated in Fig.~\ref{Q12}(Left).
In the case $\frac{3}{2} < \lambda < 2 $ the spin spiral pitch stays zero until 
the critical concentration 
$p_c=\frac{1}{\lambda}\left(1+\sqrt{1-\frac{\lambda}{2}}\right)p_0$,
and then it jumps to the single layer value (\ref{Q1}), see Fig.~\ref{Q12}(Right).
I would like to reiterate once more that the considered picture is valid for
both the static and the dynamic spirals. Ultimately, the spiral always becomes dynamic
at $p \geq 0.1-0.12$, see Ref.~\cite{milstein07}.
Clearly at $\frac{3}{2} < \lambda < 2 $ the jump at $p=p_c$ is the first order phase 
transition. On the other hand, $p_0$ and $p_1$ at
$1 < \lambda < \frac{3}{2} $ are Lifshitz points.

The value $\lambda\approx1.3$ has been obtained from the fit of LSCO 
data~\cite{milstein07}.
The parameter $\lambda$ cannot be influenced by the relatively weak interlayer 
coupling, therefore the same value should be used for YBCO~\cite{com}.
According to Refs.~\cite{doiron,miller} the AF order in YBCO extends up to $p=0.06 -0.07$,
so let us take $p_0 \approx 0.065$.
This is another parameter of the theory.
Having these two parameters one can predict the incommensurate wave vector in YBCO.
The prediction is shown in Fig.\ref{Q12}(Left) by the solid line
The theory agrees very well with neutron scattering data shown by red 
circles~\cite{hinkov07a,hinkov07b,hinkov04}.
Using (\ref{p0}) one finds the value of the bonding-antibonding splitting, 
$\Delta \approx 70\mbox{meV}$.
This is consistent with LDA calculations~\cite{andersen95} and with
ARPES data~\cite{borisenko06}.
In the same Fig.\ref{Q12}(Left)
the single layer theoretical $Q(p)$ is shown by the dashed line,
and the neutron scattering LSCO data~\cite{yamada98} are shown by the blue squares.

The developed theory is based on the small-$p$ expansion.
Therefore, it is not surprising that at $p > 0.12$ the experimental data 
start to deviate from the theory.
Note also, that the single layer formula (\ref{Q1}) is not applicable to LSCO at 
$p < 0.055$. The region $p < 0.055$ in LSCO corresponds to
the strong localization regime and the relevant theory
was developed in Ref.~\cite{luscher07}.

The incommensurate spin ordering is a generic property of underdoped
cuprates. However, the ordering properties and the phase diagrams
of the single-layer LSCO and of the double-layer YBCO are remarkably different,
see Fig.~\ref{lsco}. It is shown that while in LSCO the intrinsic disorder to a 
large extent drives the magnetic properties, the role of disorder in YBCO is
practically insignificant while the bilayer structure is crucial.
The present analysis demonstrates also that the superconductivity is intimately
related to the incommensurate spin ordering.  In LSCO this relation is masked by the
intrinsic disorder, superconductivity is impossible in the strongly-localized
regime and therefore $p_{sc}$ is determined by percolation.
However, in YBSO the correlation between superconductivity and incommensurate spin 
ordering is clear, the critical concentration for onset of superconductivity 
practically coincides with that for onset of the incommensurate spin order.
Physical mechanisms behind this observation will be considered
elsewhere.

I am grateful to O.K.~Andersen and V.~Hinkov for important discussions.


\begin{thebibliography}{99}
\bibitem{com} The hopping matrix elements $t'$ and $t''$ in YBCO
are slightly different from that in LSCO. This can give a small
difference in the value of $\lambda$.
\end{thebibliography}
\end{document}